\documentclass[twocolumn,amssymb, amsmath, aps, showpacs, footinbib, prb]{revtex4-1}

\usepackage{graphicx}% Include figure files
\usepackage{dcolumn}% Align table columns on decimal point
\usepackage{bm}% bold math
\usepackage{color}

\newcommand{\degree}{\ensuremath{^\circ}}
\begin{document}

%\preprint{APS/123-QED}%

\title{Electronic and Thermoelectric Properties of RuIn$_{3-x}A_{x}$ ($A$ = Sn, Zn)  }

\author{Deepa Kasinathan,$^1$ Maik Wagner,$^1$ Klaus Koepernik,$^2$ Raul Cardoso-Gil,$^1$ Yu. Grin,$^1$ and Helge Rosner,$^1$ }

\affiliation{$^1$Max-Planck-Institut f\"ur Chemische Physik fester Stoffe, 01187 Dresden, Germany}
\affiliation{$^2$ IFW Dresden, P.O. Box 270116, 01171 Dresden, Germany}

\begin{abstract}
Recently, we reported [M. Wagner {\it et al.} J. Mater. Res. {\bf 26}, 1886 (2011)]
transport measurements on the semiconducting
intermetallic system RuIn$_{3}$ and its substitution derivatives RuIn$_{3-x}A_{x}$ ($A$ = Sn, 
Zn). Higher values of the thermoelectric figure of merit ($zT$ = 0.45) compared to the
parent compound were achieved by chemical substitution. Here, using
density functional theory based calculations, we report on the 
microscopic picture behind the measured phenomenon. 
We show in detail that the 
electronic structure of the substitution 
variants of the intermetallic system RuIn$_{3-x}A_{x}$ ($A$ = Sn, 
Zn) changes in a rigid-band like fashion. This behavior 
makes possible the fine tuning of the 
substitution concentration to take advantage
of the sharp peak-like features in the density of states of the 
semiconducting parent compound. Trends in the transport properties calculated 
using the semi-classical Boltzmann transport equations within the
constant scattering time approximation are in good agreement with 
the former experimental results for RuIn$_{3-x}$Sn$_{x}$. Based on the calculated thermopower for the
$p$-doped systems, we reinvestigated the Zn-substituted derivative and obtained
ZnO-free RuIn$_{3-x}$Zn$_{x}$. The new experimental results are 
consistent with the calculated trend in thermopower and yield large $zT$ value of 0.8.  

\end{abstract}


\maketitle

\section{Introduction}
Thermoelectric materials, capable of creating electricity from 
waste heat are ideal in the search for sustainable energy resources. 
The bulk of present day efficient thermoelectric materials have one common 
feature: all are heavily doped narrow band gap semiconductors with a
carrier concentration of 10$^{19}$ to 10$^{21}$ cm$^{-3}$. 
In particular, narrow band gap materials with complex crystal structure have
been shown to possess excellent thermoelectric properties.\cite{aeppli1992,mahan1996}
In general, the efficiency of a thermoelectric material is measured
by a dimensionless parameter called the figure of merit, $zT =
S^{2}\sigma T/(\kappa_{el}+\kappa_{ph})$, where $S$ is the
thermopower (Seebeck coefficient) of the material, $T$ is temperature, $\sigma$ is the electrical conductivity,
$\kappa_{el}$ is the electronic part of the thermal conductivity and
$\kappa_{ph}$ is the lattice (phonon) contribution to the thermal conductivity. 
A $zT$ value larger than 1 is imperative for
successful applications. Since all the material-related 
parameters that determine $zT$ are interlinked, tuning each parameter
quasi-independently without adversely affecting the other is a tricky endeavor.

Many of the binary compounds formed by transition-metal atoms with group III, IV and
V elements have been shown to possess electronic band gaps. For example,
FeSi, FeSb$_{2}$ and FeGa$_{3}$ are all semiconducting with energy gaps of
about 0.06 eV, 0.04 eV and 0.45 eV respectively.\cite{jaccarino1967,bentien2007,tsujii2008}
The hybridization of the narrow $d$ bands of the transition metals
with the rather broad $p$ bands of the $p$-elements has been suggested 
to assist the formation of such band gaps.\cite{riseborough2000} In particular, such hybridization also 
produces sharp features close to the Fermi level, which in turn has been shown to be 
quite beneficial for enhanced thermoelectric properties.\cite{mahan1996,freericks2003} 
Consequently, large thermopower
values of +500 $\mu$V/K at 50 K in FeSi,\cite{sales1994} -45 mV/K at 10 K in FeSb$_{2}$,\cite{bentien2007} and -350 $\mu$V/K at 300  K in FeGa$_{3}$\cite{hadano2009} have been evidenced. 
The compound RuIn$_{3}$ is isotypic to FeGa$_{3}$ and was first reported\cite{Schubert1959} in 1959
and later confirmed by Holleck and co-workers\cite{Holleck1964} in 1964.
Early resistivity measurements reported poor metallic conductivity in this system,\cite{poettgen1995}
but later, density functional theory based calculations predicted semiconducting behavior.\cite{Haeussermann2002} 
Subsequent measurements have confirmed the semiconducting nature of RuIn$_{3}$.\cite{Haeussermann2002, bogdanov2007}
Studies on the thermoelectric properties of RuIn$_{3}$ do not exist. 
Recently, we reported the results of the transport measurements of both 
RuIn$_{3}$ and its
substitution variants RuIn$_{3-x}A_{x}$ ($A$ = Sn, Zn).\cite{wagner2011}
The substitution variants 
exhibit relatively large Seebeck coefficients in a wide temperature range along with 
reduced thermal conductivity compared to the parent semiconductor RuIn$_{3}$,
thereby incorporating them in the family of potential thermoelectric materials. 
Transport measurements on the binary RuIn$_{3}$ show semiconducting behavior with a band gap of $\approx$ 0.45 eV
for polycrystalline samples.\cite{wagner2011} 
The binary compound shows multi-band features exhibiting a large negative thermopower
S = -363 $\mu$V/K at 308 K and a positive thermopower S = +262 $\mu$V/K at 461 K.\cite{wagner2011}
This crossover behavior (electron transport at 308 K to hole transport at 461 K) is 
suppressed by chemical substitution.
The Sn-substituted samples show $n$-type behavior, while the Zn-substituted samples 
show $p$-type behavior, with a transition from semiconducting to metallic 
behavior with increasing $x$. Chemical substitution reduces the thermal 
conductivity to 50\% of the value of the binary compound, resulting in a $zT$ of 0.45 at 630 K for 
RuIn$_{2.95}$Zn$_{0.05}$; an improvement by a factor of 7 over pure binary phase.
The wide chemical versatility of this material, combined with the observance of 
good thermoelectric properties, suggests that further
investigations are necessary to optimize and tune this material and identify similar 
versatile members (abundant and low-cost alternatives) as viable 
candidates for future high-temperature thermoelectric applications. 

\begin{figure}[t]
\begin{center}
\includegraphics[angle=-0,width= 6.0cm,clip]{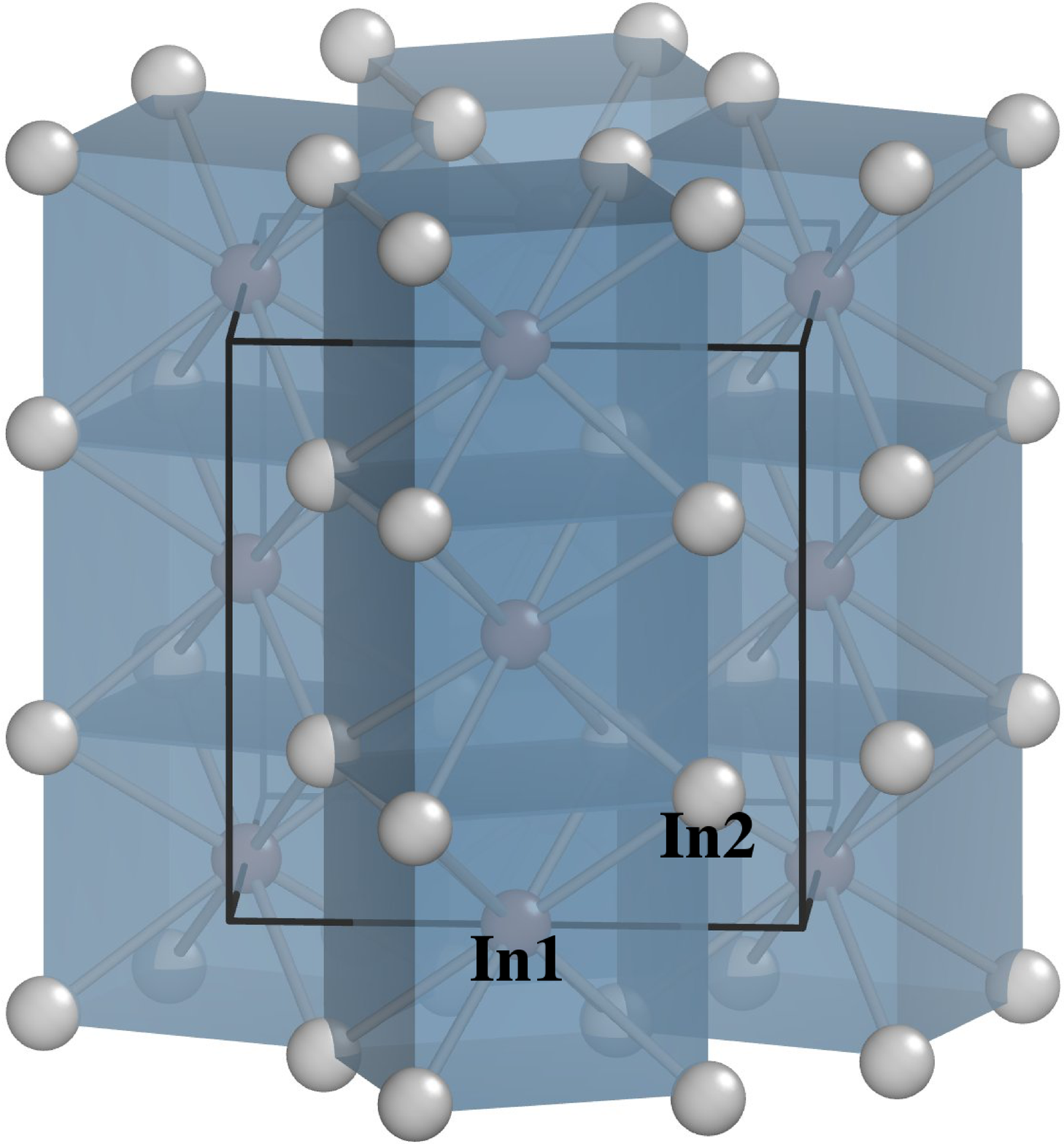}
\includegraphics[angle=-0,width=6.0cm,clip]{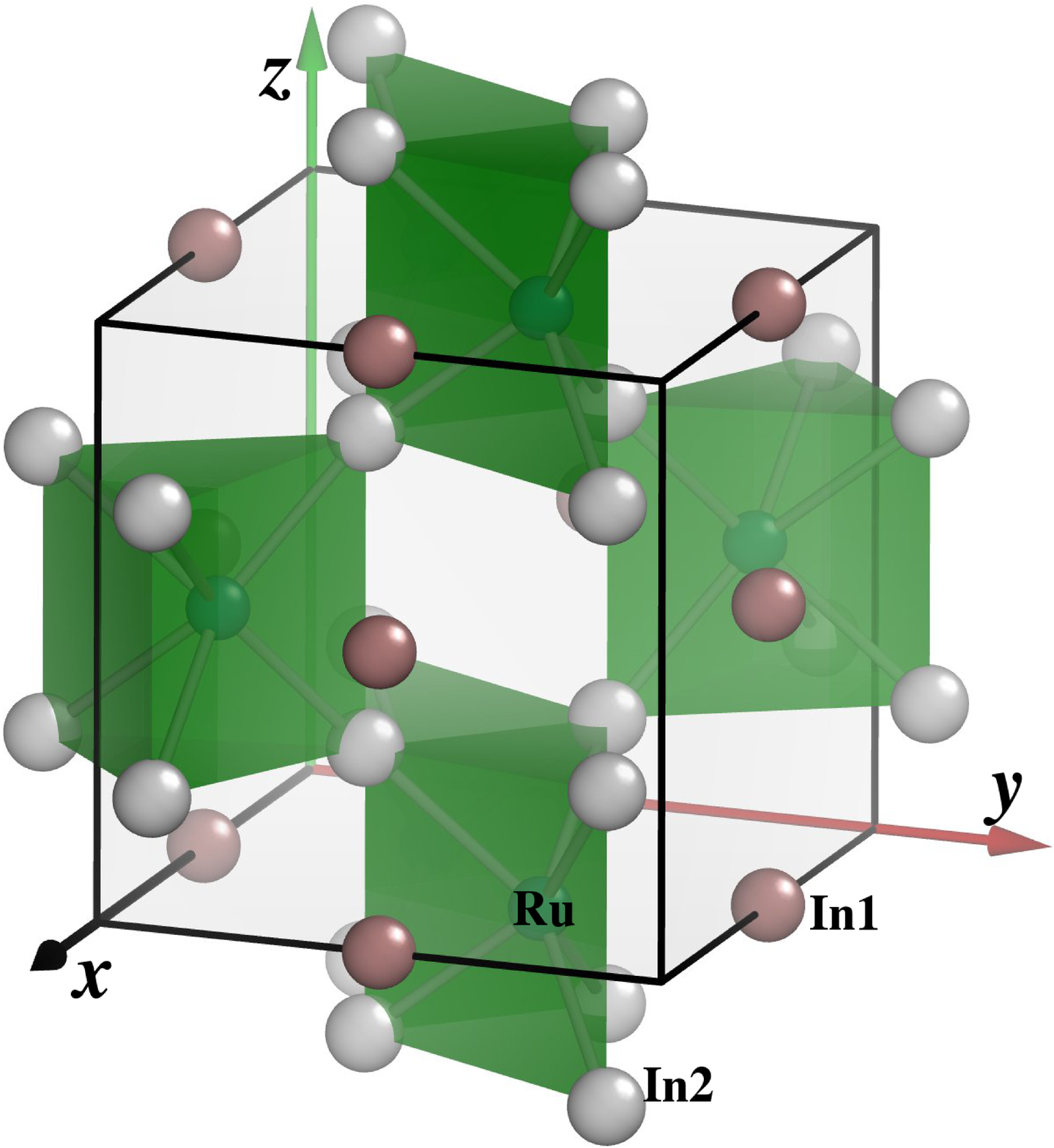}
\caption{\label{str}(Color online) Crystal structure of RuIn$_{3}$. Top panel shows the 
distorted cubes formed by In2 atoms, with In1 atoms sitting in the centers. Bottom panel
shows the distorted trigonal prisms formed by In2 atoms with Ru atoms at its centers. Two 
neighboring trigonal prisms share a rectangular face. The prisms and the cubes are interlinked 
along the $c$ axis by the corner In2 atoms. }
\end{center}
\end{figure}

Here, firstly
we present our results on the electronic and transport properties of RuIn$_{3}$
and the substitution variants using density functional theory (DFT) based calculations to
provide a microscopic insight into the physics behind the measured phenomenon.
The three theoretical reports on the parent compound that exist in literature have mainly
focussed on evaluating the band structure to obtain an estimate for the band gap. \cite{bogdanov2007,imai2006,mani2011}
In the present work, we focus on investigating the changes to the electronic 
structure as a function of substitution, as well as to probe the 
dependence of thermoelectric transport properties on the carrier concentration based
on first-principles calculations.  
By doping a semiconductor, one creates a slightly different material and the changes 
induced by doping manifest themselves as a tranformation of the band structure and, correspondingly, the 
electronic density of sates.
Various physical mechanisms, including electron-electron interaction, impurity-band widening,
band tailing, and screening effects  between the dopant and the host structure play a role in the
altering of the band structure.\cite{mertens1981} Additionally, the dopant can either be randomly distributed or 
enter the host structure in an ordered fashion. These varied scenarios can
have a profound effect on the electronic structure of the substituted variants, and 
the resulting band structure could either be a simple rigid-band shift of the semiconducting
host compound to account for the additional holes/electrons or a complex scenario with
unequal band shifts of the intrinsic band edges.
Tuning thermoelectric properties of the parent semiconductor by doping, requires a 
well controlled change in the density of states of the doped variants, thereby accessing the sharp
peak-like features close to the Fermi level of the host material. 
Consequently, it necessitates a careful study of the electronic structure of both the
doped and un-doped compounds first and the subsequent calculation of the transport properties. 
In our work, we follow this two-step approach for RuIn$_{3-x}A_{x}$ ($A$ = Sn, Zn).

In our previous work, we observed small amounts of a ZnO impurity phase and non-reacted elemental 
Ru in the RuIn$_{3-x}$Zn$_{x}$ samples.\cite{wagner2011} 
The presence of these impurity phases may result in the somewhat non-uniform behavior of 
the resistivity and thermopower values (as a function of $x$)
for RuIn$_{3-x}$Zn$_{x}$ in the temperature range 300\,K $\leq$ $T$ $\geq$ 600\,K.
On the contrary, thermopower of the Sn-substituted samples reduced uniformly with increasing $x$. 
Based on our calculations of the transport properties of these materials (section V), we infer that the thermopower of both
$n$- and $p$-doped systems should show a uniform behaviour. 
Consequently, we have successfully prepared a new batch of ZnO free RuIn$_{3-x}$Zn$_{x}$ ($x$ = 0.025,
0.050 and 0.100) applying starting materials of higher purity and measured their
transport properties.

\section{Crystal Structure}
Subsequent to the initial refinement of the crystal structure of RuIn$_{3}$  within the space group
$P\bar{4}n2$ (CoGa$_{3}$ type), it has been shown that the this compound 
crystallizes in the higher symmetry of the tetragonal space group $P4_{2}/mnm$ (no. 136), 
and belongs to 
the FeGa$_{3}$ type of structure.\cite{poettgen1995}
The lattice parameters used in our calculations are $a$ = 6.999\,\AA\, and $c$ = 7.246\,\AA.\cite{wagner2011} 
The unit cell of RuIn$_{3}$ contains four formula units, with 
two crystallographically inequivalent In sites, In1 and In2
(Ru: [0.3451, 0.3451, 0]; In1: [0, 0.5, 0]; In2: [0.1555, 0.1555, 0.2622]).\cite{poettgen1995}
The structure of RuIn$_{3}$ can be visualized using two basic
building blocks in a complementary fashion: 
(i) each In1 atom sits in the center of
a distorted cube formed by eight In2 atoms (Fig.\,\ref{str}, top panel);
(ii) around each Ru atom six neighboring In2 atoms form a distorted trigonal prism, 
with two trigonal prisms sharing a common face.
Additionally, the trigonal prisms are interlinked along the crystallographic
$z$ direction by sharing In2 corners (Fig.\,\ref{str}, bottom panel). 
Both building blocks share side faces and form a complex
three-dimensional packing of polyhedra, rather than a layered one. 

\begin{figure}[t]
\begin{center}
\includegraphics[angle=-0,width= 8.5cm,clip]{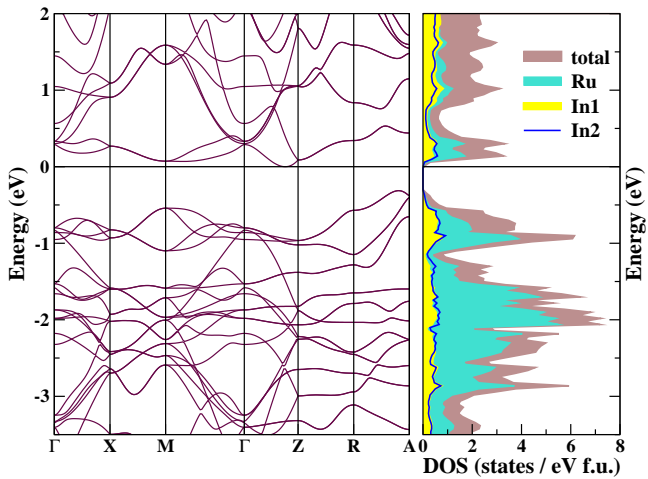}
\caption{\label{bands}(Color online) Band structure (left panel) and total and site
projected electronic density of states (right panel) of non-spin-polarized RuIn$_{3}$.
The band structure is plotted along the standard high-symmetry directions (in units of ($\pi/a,\pi/b,\pi,c$)) of
a tetragonal lattice: $\Gamma$(0,0,0) $\Longrightarrow$ X(0,1/2,0) $\Longrightarrow$
M(1/2,1/2,0) $\Longrightarrow$ $\Gamma$ $\Longrightarrow$ Z(0,0,1/2) $\Longrightarrow$ R(0,1/2,1/2)
$\Longrightarrow$ A(1/2,1/2,1/2). }
\end{center}
\end{figure}

\section{Calculational details}
Non-spin-polarized 
DFT total energy and Kohn-Sham band-structure calculations were 
performed applying the full-potential local-orbital code (version FPLO9.01.35),
within the local 
density approximation (LDA).\cite{fplo1,fplo2}
The Perdew and Wang flavor\cite{PW92} of the
exchange correlation potential was chosen for the scalar relativistic calculations.\cite{footnote1}
The calculations were carefully converged with respect to the number
of $k$ points, and finally a 24$\times$24$\times$24 
regular $k$-point mesh (1183 points in the irreducible Brillouin zone (BZ)) 
was used in the full BZ. 
The lattice parameters for undoped RuIn$_{3}$ and as well as the doped variants\cite{footnote} are taken from Ref.\,\onlinecite{wagner2011}.
It is well known that LDA tends to overbind, resulting in lattice constants that are too small. Therefore,
the calculations were carried out with experimental lattice constants, while
the internal coordinates are relaxed for all the structures considered in this work.
The substitutional derivatives RuIn$_{3-x}$Sn$_{x}$ and RuIn$_{3-x}$Zn$_{x}$ were
modeled using three kinds of approaches: the simpler virtual crystal approximation (VCA),
the fully disordered coherent potential approximation (CPA, calculated using FPLO5.00)\cite{cpa}, 
and the ordered supercell (SC) approach. Within VCA, a virtual atom with the desired number of 
electrons and protons is substituted at the indium wyckoff positions.
We employed the VCA in the following
way: the input of our full potential method are the nuclear
positions and charges and the basis functions (local orbitals). A virtual atom is
constructed such that the number of effective valence electrons equals
the one in the disordered alloy approximated by VCA. This is achieved by
choosing an appropriate non-integer nuclear charge and a corresponding
equal number of electrons for the disordered sites. The full potential
scheme then solves the DFT equations for this unit cell with non-integer
nuclear/electron charges, automatically adapting the basis functions and
the potential to this nuclear configuration as it does for any other
arrangement of "natural" atoms. This choice of VCA reproduces the
correct electron filling and often gives very satisfactory results
compared to supercell and/or CPA calculations.\cite{NJP2009,miriam}
In CPA, the disorder of the dopant is taken into account explicitly and hence 
more reliable, but also more time consuming.
Within CPA, the dopant atom retains its identity, but produces 
an effective medium that accounts for random disorder, resulting in the 
incoherence of the bands or more precisely, the spectral function $A$($k$,$\omega$). 
The incoherence arises because of the quantum 
mechanical averaging of the wavefunctions over the atomic potentials. 
In the SC approach, one of the indium position is occupied by the substituent Zn or Sn. 
Such a construction results in ordering of the substituent, but allows one to quantify 
specific effects that arise because of the periodicity that is imposed. 
Two supercells were constructed. The first was constructed by substitution of 
one In by Zn(Sn) within the tetragonal $P4_{2}/mnm$ structure unit cell of RuIn$_{3}$.
Since one unit cell contains four formula units, this corresponds to 8.33\%   
Zn(Sn) substitution and has no Zn-Zn(Sn-Sn) nearest-neighbor pairs. 
The second cell is a doubled cell along the $c$ axis with 8 formula units per 
unit cell. Again, one In was replaced by Zn(Sn), corresponding to 4.2\% Zn(Sn)
substitution. Like the smaller supercell, this cell also has no nearest-neighbor
Zn-Zn(Sn-Sn) bonds.

The  transport properties were calculated using the semiclassical Boltzmann transport 
theory\cite{semi1, semi2, semi3} within the constant scattering approximation as implemented in BoltzTraP.\cite{boltztrap}
This  approximation is based on the assumption that the scattering 
time $\tau$ determining the electrical conductivity does not vary strongly with energy on the 
scale of $kT$. Additionally, no further assumptions are made on the dependence of $\tau$ due
to strong doping and temperature. 
This method has been successfully applied to many narrow band gap materials 
including clathrates and as well as to oxides.\cite{semi3,johnsen2006,kasinathan2007,xiang2007}

\begin{figure}[t]
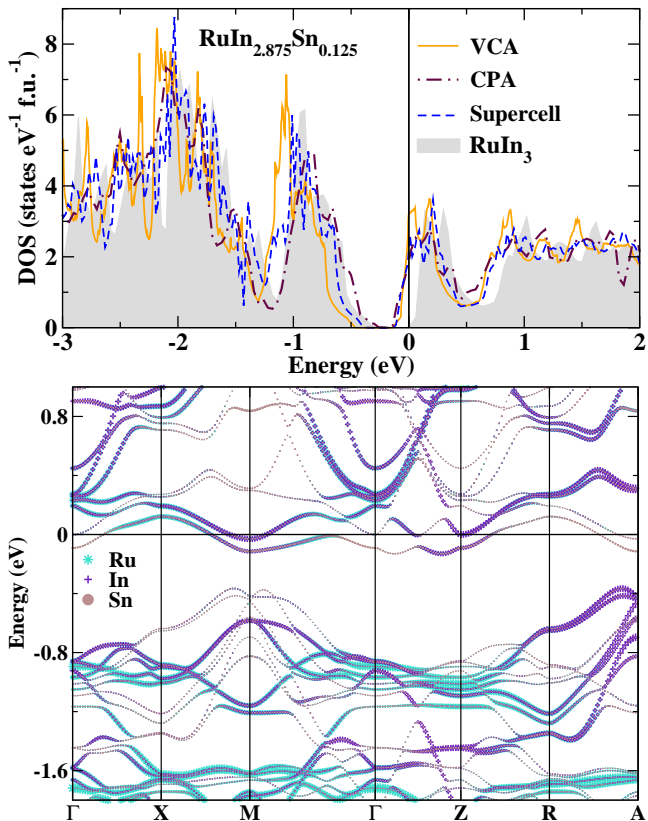

\begin{center}
\includegraphics[angle=-0,width= 8.5cm,clip]{Sndoping.eps}
\includegraphics[angle=-0, width=8.5cm,clip]{8j_Sn_bandweights.eps}
\caption{\label{sndoping}(Color online) Top panel: Calculated electronic DOS for the substitution
derivative RuIn$_{2.875}$Sn$_{0.125}$ and the binary compound RuIn$_{3}$ as a reference.
The vertical line at zero energy denotes the Fermi level and has been set to the bottom
of the conduction band of the parent compound for easy visualization of the changes upon electron doping.
The features close to the Fermi level are quite similar for VCA, CPA and the SC approach and
change the DOS in a rigid-band like fashion compared to the parent compound.
The difference between VCA and SC approach below -0.3 eV arise from the
fact that, within SC approach the introduction of an actual Sn atom with a larger extent
of the 5$p$ orbitals increases the 4$d$-5$p$ band width. Disorder induced broadening is
also observed for the CPA DOS.
Bottom panel: The non-magnetic site projected band structure of RuIn$_{2.875}$Sn$_{0.125}$ using the
SC approach. For easy comparison with the band structure of the binary compound in Fig.\,\ref{bands},
the SC approach bands have been unfolded into the smaller tetragonal unit cell. The bands crossing
the Fermi level have both In and Sn character. }
\end{center}
\end{figure}

\section{Electronic structure}
\subsection{RuIn$_{3}$}

Collected in Fig.\,\ref{bands} are the LDA non-spin-polarized band structure
and density of states (DOS) of the tetragonal RuIn$_{3}$. 
In the energy range displayed (-3.5 eV to 2 eV) there is a 
strong hybridization between the In 5$p$ states and the Ru 4$d$
states. We obtain a band gap value of 0.3\,eV, consistent with 
previously published results.\cite{imai2006,bogdanov2007,mani2011} It is well known that 
the Kohn-Sham bands obtained from contemporary exchange correlation approximations 
underestimate the size of the semiconducting band gaps, hence 
the calculated band gap of 0.3\,eV is naturally slightly smaller than the experimentally
observed value of 0.45\,eV.\cite{bogdanov2007} 
From the band structure, we witness that RuIn$_{3}$ has an 
indirect band gap with the top of the valence band occurring close
to the A point along the R$\Longrightarrow$A symmetry line and the
bottom of the conduction band is close to the Z point along the $\Gamma$
$\Longrightarrow$Z line. It is worthwhile to note that the curvature of 
these two bands are
similarly parabolic, resulting in similar values for the effective masses for 
the holes and electrons at low substitution concentrations. Our results concerning the bottom of the 
conduction band are consistent with the reports of Imai and co-workers\cite{imai2006} and
with that of Mani and co-workers.\cite{mani2011} On the contrary, calculations by Bogdanov {\it et al.}\cite{bogdanov2007} 
using the TB-LMTO-ASA (tight binding-linearized muffin tin orbital-atomic sphere approximation)
package find the rather flat band along M$\Longrightarrow\Gamma$ as the minimum
of the conduction band thereby yielding a larger effective mass of the 
holes as compared to electrons.
Additionally, the bands are similarly dispersive along the various 
high symmetry directions, consistent with the three dimensional nature
of the building blocks in the crystal structure. 
Our calculations are well in agreement with the recent experimental results on RuIn$_{3}$.\cite{wagner2011}

\subsection{RuIn$_{3-x}$Sn$_{x}$: Electron doping}

The consequences of substituting Sn (group 
IV) for In predicted by the three calculational approaches are 
displayed in Fig.\,\ref{sndoping}. The top panel compares the
total DOS
for the various approaches for the 4.2\% substitution scenario RuIn$_{2.875}$Sn$_{0.125}$.
Compared to In, Sn has one additional electron and 
acts as an electron donor in the substitution variant. For easy comparison, the DOS of the
parent compound is also included in the figure.
RuIn$_{2.875}$Sn$_{0.125}$ is metallic compared to the parent compound which is 
semiconducting (see Fig.\,\ref{bands}). 
The shape of the DOS close to the Fermi edge is 
quite similar between VCA, CPA and the SC approach for RuIn$_{2.875}$Sn$_{0.125}$.
The slight dissimilarity in the DOS away from the Fermi edge arise from the underlying differences in the 
methodologies.
Explicit inclusion (as compared to VCA) of the Sn atom with the more
extended  5$p$ orbitals in the ordered SC approach results in 
a larger band width of the Ru 4$d$ - In 5$p$ - Sn 5$p$ hybridized states. This is clearly
discernible in the DOS below -0.3 eV. Similarly, disorder induced incoherence of the valence band 
is also noticed for the CPA derived DOS below -0.3 eV.
As mentioned previously, there are two crystallographically inequivalent indium positions
in this structure (In1 at $4c$ and In2 at $8j$). Hence, for each SC, two different
calculations were performed by placing the substituent Sn once at the $4c$ position and once
at the $8j$ position. 
The Sn atom preferentially occupies the 8$j$ (In2) position, which is energetically favorable by
20 meV per formula unit compared to the 4$c$ (In1) position. 
The bottom panel of Fig.\,\ref{sndoping} shows the non-magnetic band structure of RuIn$_{2.875}$Sn$_{0.125}$
obtained using the SC approach. For a 4.2\% substitution, the original unit cell of the 
binary compound has been doubled along the $z$ axis, resulting in a smaller Brillouin zone (BZ). 
This has two consequences, firstly there are twice the number of  bands and secondly all the 
bands are folded back into the small BZ of the SC. 
The resulting band picture gets quite complicated and hard to 
discern meaningful information from it.
To facilitate easy comparison with the band structure of the binary compound, we have unfolded the bands according to
the translational symmetry of the original unit cell using the approach described in Refs.\,\onlinecite{heumen2011,ku2010}.
The Ru and In character unfolded bands of the SC in Fig.\,\ref{sndoping} are comparable to the bands in Fig.\,\ref{bands},
though most of the degeneracies are lifted due to the presence of the Sn substituent. 
The size of the symbols in the band structure plot refers to the $k$-resolved
weight of all the orbitals of the chosen site. The larger the size of the
symbol, the larger is the contribution of the site to the band structure. 
Based on this, we can conclude that 
the bands crossing the Fermi level have both In and Sn character, comprehensible due to the
fact that Sn acts as an electron donor, making the system metallic and is only one atomic number away from In in the
periodic table, and hence similar to In. 

\begin{figure}[t]
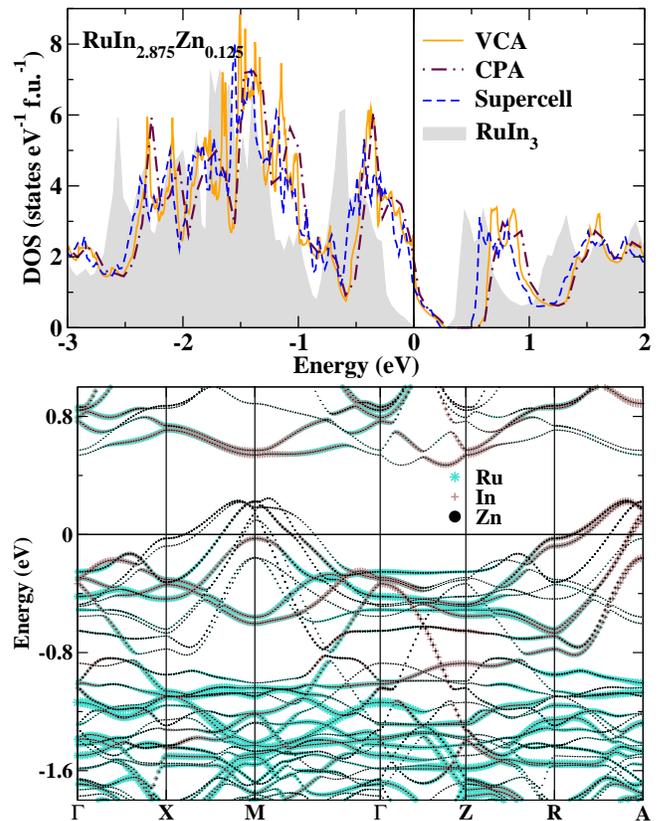

\begin{center}
\includegraphics[angle=-0,width= 8.5cm,clip]{Zndoping.eps}
\includegraphics[angle=-0, width=8.5cm,clip]{4c_Zn_bandweights.eps}
\caption{\label{zndoping}(Color online) Top panel: Calculated electronic DOS for the substitution
derivative RuIn$_{2.875}$Zn$_{0.125}$ and the binary RuIn$_{3}$ for reference.
The vertical line at zero energy denotes the Fermi level and has been set at the top of
the valence band of the parent compound for easy visualization of the changes upon hole doping.
VCA, CPA and the SC approach result in a metallic ground state with similar features of the DOS close to the
Fermi edge. Analogous to the electron doping results, the changes in DOS for the hole
doped system are again rigid-band like.
In contrast to Sn doping (see caption of Fig.\,\ref{sndoping}), Zn 4$d$ orbitals are not
that extended and hence the valence band spectrum is quite similar for both VCA and
SC approaches.
Bottom panel: The non-magnetic site projected band structure of RuIn$_{2.875}$Zn$_{0.125}$ using the
SC approach. For easy comparison with the band structure of the parent compound in Fig.\,\ref{bands},
the SC approach bands have been unfolded into the smaller tetragonal unit cell. The bands crossing
the Fermi level have a strong In character. }
\end{center}
\end{figure}

Comparing the DOS of the binary RuIn$_{3}$ with that of RuIn$_{2.875}$Sn$_{0.125}$,
we observe a rather rigid-band-like shift in the electronic structure.
Similar behavior is observed for other substitution concentrations (not shown here). 
This result is extremely interesting from a technological point of view. 
Mahan and Sofo\cite{mahan1996} showed that a narrow distribution of the 
energy of the electrons participating in the transport process is needed to maximize 
thermoelectric efficiency. 
The binary RuIn$_{3}$ possesses sharp peak-like features close to the
conduction band minimum and as well as near the valence band maximum. 
Electron doping, as shown here using Sn as a substituent, results in a rigid-band
shift of the DOS without adversely affecting the sharp peak-like features, thereby
opening the possibility of fine-tuning the substitution to achieve maximum thermoelectric 
efficiency. This will be discussed in detail in section V.

\subsection{RuIn$_{3-x}$Zn$_{x}$: Hole doping}

Following the same steps as described above for the Sn substitution, we have calculated the 
electronic structure of the Zn substituted system as well. Collected in Fig.\,\ref{zndoping} are the
DOS and unfolded band structure for the 4.2\% substitution scenario RuIn$_{2.875}$Zn$_{0.125}$. 
Substitution of Zn introduces holes into the system making RuIn$_{2.875}$Zn$_{0.125}$
metallic compared to the binary RuIn$_{3}$. 
Similar to the observations in the electron doped system, the DOS close to the Fermi edge
in the hole doped variant is comparable between VCA, CPA and the SC approach.
Moreover, the 3$d$ orbitals of Zn are 
less extended and fully filled, resulting in a strong In character of the bands that cross the
Fermi level. 
Nevertheless, the changes in the DOS close to the Fermi level are still rigid-band 
like, beneficial in tuning transport properties. 
Contrary to the Sn doping scenario, Zn atoms desirably occupy the 4$c$ (In1) position which
is energetically favorable by 27 meV per formula unit compared to the 8$j$ (In2) position. 
This difference in preferred site occupations between the two substituents can be 
construed as follows:
In2 atoms form a distorted trigonal prism with a Ru atom at its center (refer Fig.\,\ref{str}).
Extended $p$ orbitals of the anions are necessary to facilitate the hybridization
with the narrow $d$ bands of the transition metal cations at the center of the distorted 
trigonal prism. Hence, Sn atoms which possess extended and partially filled $p$ orbitals preferentially
occupy the In2 position.
Contrariwise, the transition metal Zn atom lacks extended $p$-like orbitals 
and hence preferentially occupies the In1 position, which is at the center of a distorted cube made of 8 
neighboring In2 atoms.
Our theoretical observation of preferential site occupation for the Zn atom is 
consistent with the experimental observation in another isotypic system 
CoIn$_{3-x}$Zn$_{x}$.\cite{viklund2002}
Using X-ray diffraction and neutron powder diffraction experiments, the 
authors observed that the substitution of In by Zn takes place in 
an ordered fashion, producing colored variants of the parent compound CoIn$_{3}$ with
the Zn atom entering exclusively the position corresponding to the cube centers. 

\begin{figure}[t]
\begin{center}
\includegraphics[angle=-0,width=8.5cm,clip]{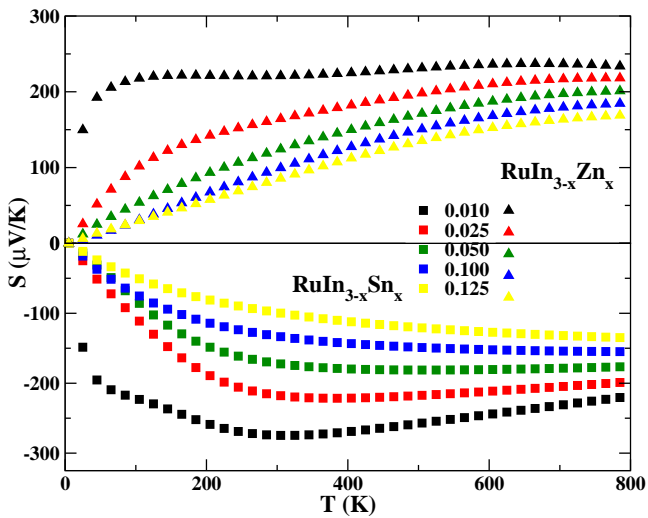}
\caption{\label{seebeck}(Color online) Thermopower S as a function of temperature for both Zn and Sn substitution
variants RuIn$_{3-x}A_{x}$. A negative sign in S represents $n$-type behavior
and a positive sign in S represents $p$-type behavior. This figure is
directly comparable to the experimental
data from Ref.\,\onlinecite{wagner2011}.  }
\end{center}
\end{figure}

At this juncture, it should be noted that the proclivity of the substituent to certain 
indium sites is discerned only from the differences in the total energies of the various 
supercell configurations. No strongly distinguishable relevant features are noticeable in 
the DOS plots when Sn(Zn) occupy either In1 or In2 position and a similar rigid-band like
shift of the DOS compared to the binary RuIn$_{3}$ is noticed for all considered configurations. 
This observation is relevant for the experiments, where the fine tuning of the
doped samples can be performed without the fear of adversely affecting the 
electronic structure of the final compound. 
Additionally, the metallic radii of In and Sn are similar ($\approx$1.60\,\AA) and hence  
size effects do not play a significant role for the electron doped systems.
On the other hand, the metallic radius of Zn is about 13\% smaller ($\approx$1.40\,\AA) than that of In
and  hence 
Zn is incorporated easily as a substitute for In in RuIn$_{3}$. 
Owing to the smaller metallic radius of the dopant Zn, there is a likelihood for local structural relaxation around the
dopant which could have a detrimental effect on the thermoelectric transport by altering the electronic 
properties of the host. 
We have investigated this in detail, by   
allowing for the local relaxation around the Zn atom in RuIn$_{3-x}$Zn$_{x}$  with $x$ = 0.125. 
In the un-relaxed structure, the Zn atom (occupying the In1 position) is surrounded by a distorted cube
of In2 atoms (see Fig.\ref{str}) with four In2 neighbors at 3.16\,\AA \, and another four In2 neighbors
at 3.26\,\AA.  The Ru-Zn bond length is 2.65\,\AA.
The structural relaxation reduces the Zn-In2 bond length by just 2\,\%, while the
Ru-Zn bond length on the other hand is slightly increased to 2.70\,\AA. 
Nevertheless, these changes did not alter the electronic structure 
significantly, likely 
since the
trigonal prisms centered around Ru and the cubes centered around In1 site are heavily
distorted, already in the undoped system.

\section{Thermoelectric properties}

\subsection{Calculations}

\begin{figure}[t]
\begin{center}
\includegraphics[angle=-0,width=8.5cm,clip]{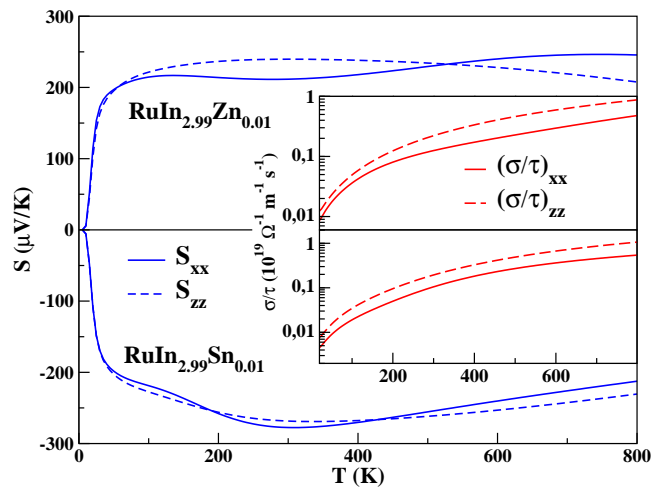}
\caption{\label{seebeck_axis}(Color online) The main panel shows the calculated thermopower
$S$ and the inset shows the electrical conductivity ($\sigma$) relative to relaxation time ($\tau$)
resolved along the crystallographic $a$
($S_{xx}$, ($\sigma/\tau$)$_{xx}$) and $c$ (S$_{zz}$, ($\sigma/\tau$)$_{zz}$) directions.    }
\end{center}
\end{figure}

Having an improved understanding of the electronic properties of 
RuIn$_{3-x}A_{x}$ ($A$ = Sn, Zn) systems, we now calculate the 
transport coefficients using the semi-classical Boltzmann theory and
the rigid-band approach. Such a procedure has been shown to successfully
predict optimal doping levels in other thermoelectric materials, for example 
clathrates\cite{semi3} and Sb$_{2}$Te$_{3}$.\cite{thonhauser2003}
One main concern when calculating transport coefficients, is the underestimation
of band gaps using the standard DFT functionals. Such an underestimation of
band gaps manifests in the reduction of thermopower at higher temperatures due
to bipolar conduction. To overcome this problem, we concentrate on analyzing the trend in 
the calculated thermopower for various concentrations of electron and
hole doping rather than quantifying them. This approach results in a consistent scenario
between experiment and theory as shown below. 
Firstly, as seen in Fig.\,\ref{bands}, it is important to note that the valence and 
conduction band of the stoichiometric compound are parabolic. As shown above,
both Sn and Zn substitution act in a rigid-band like fashion. Therefore, 
for small amounts of
substituents, the bands will continue to remain parabolic. Hence, experimentally the transport 
can be modeled using simple parabolic band expressions to obtain the concentration
of charge carriers in the system. 
We also notice that close to the conduction band minimum, there are multiple
heavy bands ({\it i.e.} bands with less dispersion and hence larger 
effective mass), which is advantageous for thermoelectric transport. 
In contrast, the valence band maximum has a larger energy difference to 
the next available heavy band. 
Consequently, from just the view point of electronic structure, an $n$-type material would retain better thermoelectric 
transport coefficients than a $p$-type for low doping concentrations and low temperatures. 

\begin{figure}[t]
\begin{center}
\includegraphics[angle=-0,width=8.5cm,clip]{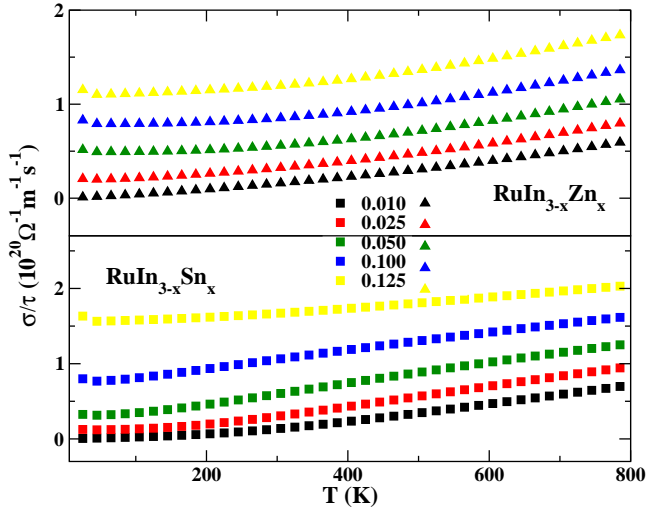}
\caption{\label{conductivity}(Color online) Calculated electrical
conductivities ($\sigma$) relative to relaxation time ($\tau$)
as a function of temperature for both Zn and Sn substitution
variants of RuIn$_{3-x}A_{x}$.    }
\end{center}
\end{figure}

\begin{figure}[t]
\begin{center}
\includegraphics[width=8.0cm]{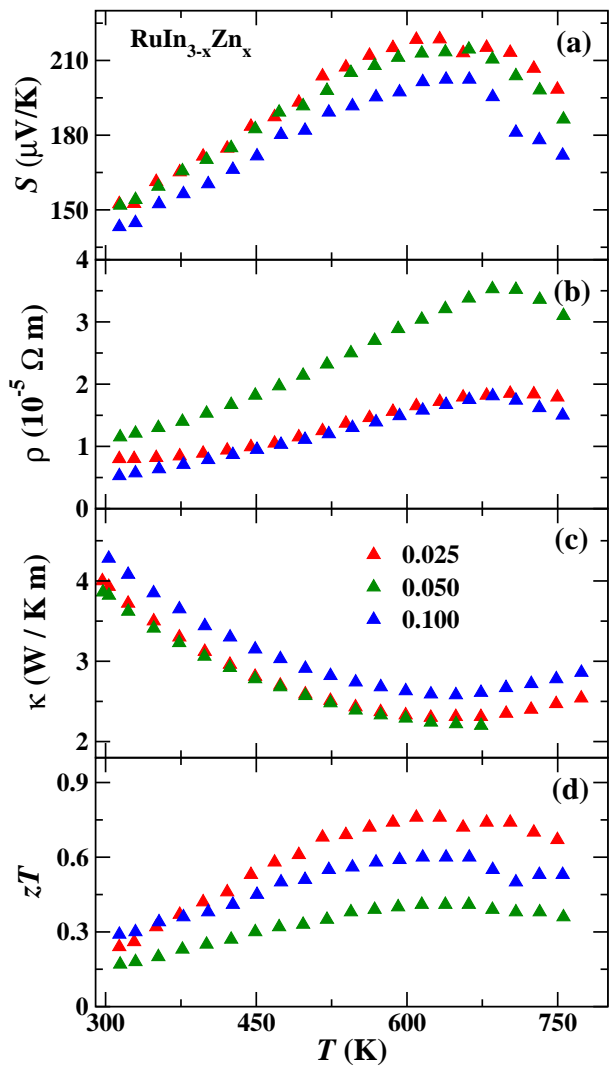}
\caption{\label{expt}Measured thermoelectric properties of RuIn$_{3-x}$Zn$_{x}$ as a 
function of temperature. (a) Seebeck coefficient $S$, (b) electrical resistivity $\rho$, (c) total thermal conductivity $\kappa$, 
and (d) dimensional figure of merit $zT$.    }
\end{center}
\end{figure}

The values of the calculated thermopower are not affected by
the constant scattering time approximation used to calculate the various
transport properties, since the expression for $S$ is independent of 
$\tau$. This means that $S$ is directly dependent on the electronic 
structure of the material. 
Collected in Fig.\,\ref{seebeck} is the temperature dependence of thermopower 
for various substitution concentrations of Zn and Sn in RuIn$_{3}$.  
Sn substitution introduces additional electrons into the system resulting in negative
values of thermopower. Correspondingly, we obtain positive values of thermopower
for Zn substitution which introduces holes in the system. 
Due to the lack of Hall data on all the measured samples, 
the assignment of the substitution concentration for the plots here are done 
by comparing the experimental thermopower measurements with the
calculated values. 
For all substitution concentrations considered here, the magnitude of $S$ increases
steeply with temperature until 300 K. Above 300 K, in RuIn$_{3-x}$Sn$_{x}$ $\lvert S \rvert$
slightly reduces or remains constant. On the other hand, for RuIn$_{3-x}$Zn$_{x}$
$\lvert S \rvert$ above 300 K continues to increase with a smaller slope. 
Below room temperature and for $x$ $\leq$ 0.050, $\lvert S \rvert$ for the
electron doped system is 1.5 times that of the hole doped system, consistent
with the electronic band structure. 
In contrast, with increased doping ($x >$ 0.050) and temperatures (T $>$ 300 K), hole doped RuIn$_{3-x}$Zn$_{x}$ 
shows increased thermopower compared to the electron doped RuIn$_{3-x}$Sn$_{x}$. 
The maximum values of \textbar $S$\textbar (larger than 200 $\mu$V/K) are obtained for the smallest doping concentrations
in both $p$- and $n$-type. 
Our results, including the absolute values of $S$,
as well as the trend upon increased doping for the $n$-type, are consistent
with the recent experimental measurements.\cite{wagner2011}
A direct comparison of calculations for the $p$-type with the recently published experiments 
is inhibited by the presence of impurity phase ZnO in the measured samples and hence 
no clear trend is discernible from the experiments.\cite{wagner2011}
Presently, stimulated by our calculational results,we have successfully synthesized a new batch of ZnO free RuIn$_{3-x}$Zn$_{x}$ 
samples and measured their thermoelectric properties (details described in section VI).
The trend in measured $S$ is now consistent with our calculated results for 
the $p$-doped systems too (refer to Fig.\,\ref{expt}(a)). 

\begin{figure}[t]
\begin{center}
\includegraphics[angle=-0,width=8.5cm,clip]{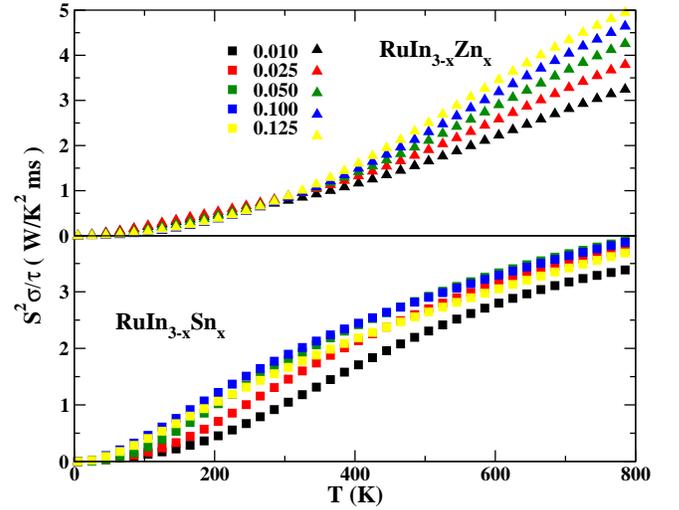}
\caption{\label{powerfactor}(Color online) Calculated power
factor S$^{2}\sigma$ with respect to relaxation time
$\tau$ as a function of temperature for both Zn and Sn substitution
variants of RuIn$_{3-x}A_{x}$.    }
\end{center}
\end{figure}

\begin{figure*}[t]
\begin{center}
\includegraphics[height=9cm,clip]{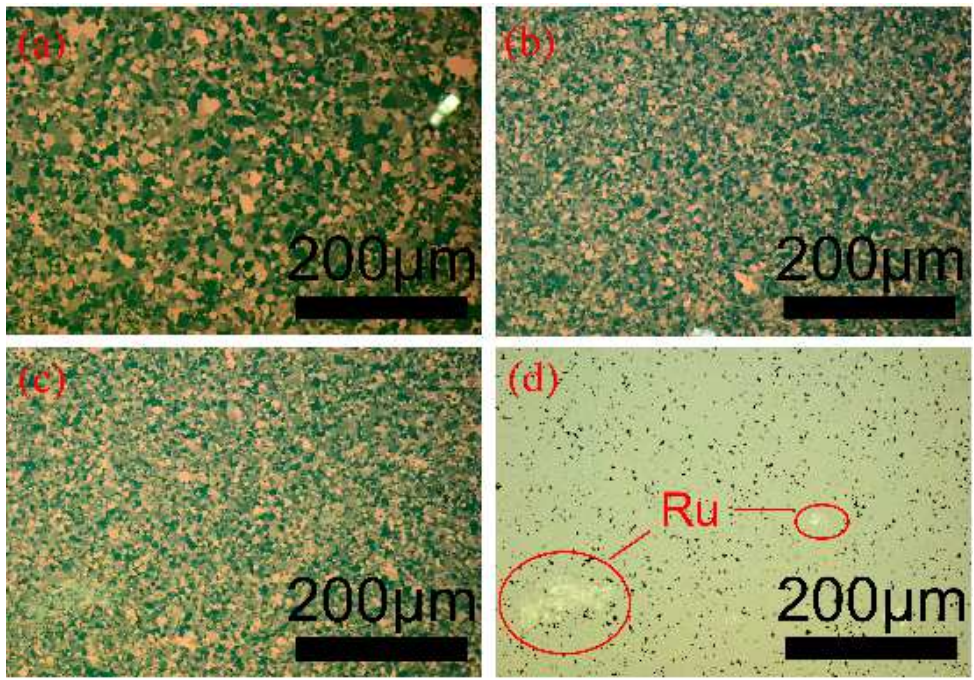}
\caption{\label{microstr}(Color online) 
Microstructure of RuIn$_{3-x}$Zn$_{x}$ samples: (a) $x$ = 0.025, polarized light; (b) $x$ = 0.050, polarized light,
the surface is perpendicular to the pressure direction during the SPS treatment; (c) $x$ = 0.050, polarized light, the surface
is parallel to the pressure direction during the SPS treatment; (d) $x$ = 0.050, bright light. }
\end{center}
\end{figure*}

Furthermore, transport measurements on the parent compound RuIn$_{3}$ find
a change in the sign of $S$ from $n$-type to $p$-type around 350 K.
Similar behavior was observed for the related compound RuGa$_{3}$.\cite{amagai2004}
The authors assigned the sign change in $S$ to the shift
of the valence band to an extrinsic region along with
the presence of light holes in the
system and proposed a two-band model, one each for electron and
hole, respectively for evaluating $S$. In this scenario, holes with 
higher mobilities can influence $S$ and change the sign from negative to
positive values.\cite{amagai2004} 
However, modeling transport properties including the presence of extrinsic 
charge carriers within a bulk system is 
more involved and beyond the scope of the present investigation.  

As discussed earlier, notwithstanding the tetragonal symmetry of 
the crystal structure of RuIn$_{3}$, the polyhedra of the basic building blocks 
form a three dimensional packing. 
Previously, measurements on single crystals of RuIn$_{3}$ have 
found a somewhat weak anisotropic resistivity along [110]
and [001] orientations.\cite{bogdanov2007}
Collected in Fig.\,\ref{seebeck_axis} are the 
values of $S$ and $\sigma/\tau$ along the crystallographic $a$ and $c$ directions.
Consistent with the three dimensional packing of polyhedra,  the
anisotropy is moderate in the calculated transport coefficients, with the 
$p$-doped system displaying a larger anisotropy than that of the
$n$-doped variant. 
This result can be qualitatively understood by analyzing 
the band structures shown in Figs.\,\ref{sndoping} and \ref{zndoping}.
For the $n$-doped system, the bands crossing the Fermi level 
disperse in a similar fashion along the various symmetry directions.
On the contrary, for the $p$-doped system, the bands crossing
the Fermi level are more dispersive in the $a-b$ plane
(X$\Longrightarrow$M$\Longrightarrow\Gamma$) compared to the 
$c$ diretion ($\Gamma\Longrightarrow$Z). 

Collected in Fig.\,\ref{conductivity} are the calculated electrical 
conductivities $\sigma$ relative to the relaxation time $\tau$ for the same
set of systems considered in Fig.\,\ref{seebeck}. 
Note that $\sigma$ has a linear dependence to $\tau$.
Assuming a constant $\tau$ for the various substitution variants, 
the increasing conductivity for either hole or electron doping reflects the fact that
more charge carriers are present for the transport process.
 Due to the presence of a rather flat band with a larger effective mass close to the 
 conduction band minimum, Sn substituted systems exhibit
 smaller electrical conductivities compared to the Zn substituted systems at 
 low temperatures and low substitution concentrations. 
 The trend observed for the thermopower previously is thus different than
 that of the conductivity, which will result in an
 interesting trend for the power factor ($S^{2}\sigma$), the numerator in
 the expression for the thermoelectric figure of merit $zT$.
 To this end, we have plotted $S^{2}\sigma$ relative to the relaxation time $\tau$ 
 in Fig.\,\ref{powerfactor}.
 For temperatures up to 300 K, the power factors of the Sn substituted
 systems are enhanced compared to those of the Zn substituted systems.
 Beyond 300 K, the trend is reversed and the hole doped systems 
 exhibit an enhanced power factor.
 A quantitative estimate for the dimensionless figure of merit $zT$ 
 from DFT, entails an accurate estimation of $\kappa_{ph}$, the lattice
 thermal conductivity. This involves solving the phonon 
 Boltzmann transport equations exactly and requires an accurate description of the
 harmonic and anharmonic interatomic forces, which is presently a 
 formidable task and is left for a more extended study in the future. Additionally, disorder and nano-structuring effects
 that arise during experiments modify $\kappa_{ph}$ in a not so 
 straightforward way for a concise theoretical description.
 Nevertheless, clues provided by the transport coefficients that are 
 directly dependent on the electronic structure have been used 
 to successfully tune the thermoelectric properties of other known 
 materials.\cite{semi3,thonhauser2003,viklund2002}
 In the case of RuIn$_{3-x}Sn_{x}$ and RuIn$_{3-x}$Zn$_{x}$, our calculations 
 of the power factor establish a robust scenario for both
 $p$ and $n$ doped systems with improved values for certain 
 doping levels.
 
 \subsection{Experimental data for R\lowercase{u}I\lowercase{n}$_{3-x}$Z\lowercase{n}$_{x}$}
 
The chemical analysis of the commercially available starting materials (Ru, In and Zn) showed 
a measurable amount of oxygen content only in the case of Ru: 0.54 mass \% of oxygen, which 
is presumably present in the form of RuO$_{2}$. 
This oxide impurity is completely removed via hydrogen reduction at 900 $\degree$C for 5 hours 
in a home-made set up (O$_{2}$ not detectable after H$_{2}$ reduction).
After this hydrogen treatment, the samples were prepared strictly in accord to the 
description in Ref.\,\onlinecite{wagner2011}.
Microstructure analysis of RuIn$_{2.975}$Zn$_{0.025}$ (Fig.\,\ref{microstr}a) show single-phase homogenous products without
any ZnO impurity.  Only in the case of RuIn$_{2.95}$Zn$_{0.05}$ and RuIn$_{2.90}$Zn$_{0.10}$ small trace
amounts of residual Ru were found (Fig.\,\ref{microstr}d), which might be due to the 
defect occupation of the Ru site. The grains, with respect to their size, shape and 
orientation are quite similar for all the samples discussed here. Additionally, the microstructure is
independent from the substitution concentration of Zn, and the direction of pressure during the 
spark plasma sintering (SPS) experiments (Fig.\,\ref{microstr}b and c). 
The X-ray diffraction patterns of the RuIn$_{3-x}$Zn$_{x}$ samples were indexed with the 
$P4_{2}/mnm$ (no. 136) space group. Compared to the previously published results\cite{wagner2011}
the $a$ lattice parameter remains unchanged, while the $c$ lattice parameter is slightly reduced (Table.\,I).
For composition of the samples, see Ref.\,\onlinecite{foot}.
Because of  the absence of ZnO impurities, all of the Zn has been incorporated 
into the lattice, resulting in a pronounced reduction of the unit cell in the $c$ direction.

The transport-related quantities and the resultant figure of merit $zT$ are collected in Fig.\,\ref{expt}. 
The temperature dependence of the electrical resistivity $\rho$ shows metallic behavior for all the substitution
derivatives, with increasing resistivity for increasing temperature. 
The absolute values of $\rho$ shown here are reduced in comparison to
our previously published results\cite{wagner2011} due to the absence of ZnO impurity phase, which is an 
insulator with a large band gap of more than 3 eV. 
The measured thermopower $S$ is positive ($p$-type) and is reducing with increasing $x$. 
At 350\,K, the absolute values of $S_{exp}$ change from 162 to 150 $\mu$V/K for 0.025 $\leq$ $x$ $\geq$ 0.100,
compared to our calculations where $S_{theo}$ change from 170 to 115 $\mu$V/K. 
For all the samples considered here, $S$ increases with increasing temperature and reaches a maximum 
around 600\,K.
This trend in $S$ as a function of $x$ and temperature are now consistent with our above mentioned theoretical 
predictions (Fig.\,\ref{seebeck}).
The thermal conductivity $\kappa$ is only slightly increased in comparison to Ref.\,\onlinecite{wagner2011}.
Previously, the presence of ZnO impurities could have acted as additional scatterers to reduce $\kappa$, which 
is presently removed from our samples. 
The temperature dependence of the dimensionless figure of merit $zT$ increases 
strongly, reaching a value of 0.8 at 620 K for RuIn$_{2.975}$Zn$_{0.025}$. 
Albeit the trend in $S$ is congruent between experiment and theory, the same cannot be
inferred for the other transport related quantities and hence $zT$, because of the complex depedency
relations between these parameters. 
While the expression for $S$ is free of parameters, $\rho$, $\kappa$ and hence $zT$ are dependent 
on the scattering time $\tau$. Our calculations which are based on the constant scattering time approximation
cannot be expected to yield more realistic results. Nonetheless, we have demonstrated that insights from
band structure calculations can be quite helpful in fine tuning thermoelectric properties. 

\begin{table}[t]
\begin{ruledtabular}
\begin{tabular}{l  c c }
$x$ & a (\AA) & c (\AA) \\
\hline
0.025 & 6.9988(1) & 7.2456(2) \\
0.050 & 6.9988(1) & 7.2396(1) \\
0.100 & 6.9977(2) & 7.2422(3) \\
\end{tabular}
\end{ruledtabular}
\caption{\label{table} Refined lattice parameters for RuIn$_{3-}x$Zn$_{x}$ }
\end{table}

\section{Conclusions}
We have analyzed in detail the electronic 
structure of the hole- and electron-doped RuIn$_{3}$ intermetallic
compound.
We have used various methods to model the effects of 
doping, starting from the simple VCA description to the time consuming
and the more reliable CPA approach. Effects of ordering of the dopant was
also considered using the SC approach.  
Comparing the results of VCA, CPA and the SC approach,
we can conclusively determine that the dopants introduce
charge carriers and change the electronic structure in a rigid-band-like 
fashion. This scenario allows for the advantageous use of 
the sharp peak like features in the DOS to obtain good thermoelectric
properties. Transport coefficients (thermopower and power factor) 
calculated using the semi-classical
Boltzmann transport equations,  are in good agreement with the 
experimental results. 
Based on our calculations, we can conclude that low substitution levels 
are advantageous in obtaining large values of thermopower. 
Both $n$- and $p$-doped systems have similar transport properties. 
In particular, based on our calculations, for $T \geq$ 400\,K, $p$ doped samples with $x \geq$ 0.125 
 show considerable potential for improving the figure of merit $zT$.
 Similarly, for $T \leq$ 500\,K, $n$ doped systems with an 
 optimal substitution in the range 0.050 $\leq x \geq$ 0.100 also show great thermoelectric potential.
A $zT \approx$ 0.8 has been obtained for RuIn$_{2.975}$Zn$_{0.025}$,
 while the first trial of experiments were able to only reach a $zT \approx$ 0.1 in the 
 $n$-doped samples.\cite{wagner2011} 
 Additional experimental investigation exploring higher $p$-dopant concentrations as well as 
 fine tuning the $n$-doped systems  are desirable to probe
 the possibility of further enhancing $zT$.  
RuIn$_{3-x}A_{x}$ ($A$ = Sn, Zn) systems with their robust electronic
properties, containing low-toxic combination of elements 
provide - although expensive - an ideal play ground for experimentalists to investigate 
in detail 
and tune the thermoelectric properties as a function of doping. 
Our combined results of theoretical calculations and experimental work has clearly demonstrated that narrow-band-gap semiconductors
formed by transition-metal atoms with group III, IV and V elements 
have a large potential in thermoelectric applications. The presented study will,
hopefully, also stimulate the search for new better and cheaper systems in this family of 
compounds whose properties follow the same underlying physics as 
RuIn$_{3-x}A_{x}$ ($A$ = Sn, Zn).

Acknowledgement: DK and HR acknowledge funding by the DFG within SPP 1386.


\begin{thebibliography}{20}
\bibitem{footnote1}Calculations using the generalized gradient approximation (GGA) yield a very
similar physical picture and confirm the robustness of our results.
\end{thebibliography}
\end{document}